\documentclass[preprint,showpacs,preprintnumbers,amsmath,amssymb,prl,superscriptaddress]{revtex4} 
\usepackage{graphicx}
\usepackage{dcolumn}
\usepackage{bm}
\usepackage{amsmath}
\usepackage{amssymb}
\usepackage[rightcaption]{sidecap}
\usepackage[justification=raggedright]{caption}
\usepackage{color}
\begin{document}
\newcommand{\refeqn}[1]{Eqn.~(\ref{#1})}
\newcommand{\reffig}[1]{Fig.~\ref{#1}}           
\newcommand{\reffigBEGIN}[1]{Figure ~\ref{#1}}       
\newcommand{\reffigBEGINP}[2]{Figure ~\ref{#1}(#2)}  
\newcommand{\reffigP}[2]{Fig.~\ref{#1}(#2)} 
\newcommand{\reffigPP}[2]{Fig.~\ref{#1}#2} 
\newcommand{\reffigPouter}[2]{(Fig.~\ref{#1}#2)} 

\title{Pinwheel stabilization by ocular dominance segregation}
\author{Lars Reichl}
\email{reichl@nld.ds.mpg.de}
\affiliation{Max-Planck-Institute for Dynamics and Self-Organization, G\"ottingen, Germany}
\affiliation{Bernstein Center for Computational Neuroscience, G\"ottingen, Germany}
\author{Siegrid L\"owel} 
\affiliation{Institute of General Zoology and Animal Physiology, University Jena, Jena, Germany}
\author{Fred Wolf}
\affiliation{Max-Planck-Institute for Dynamics and Self-Organization, G\"ottingen, Germany}
\affiliation{Bernstein Center for Computational Neuroscience, G\"ottingen, Germany}

\keywords{visual cortex, development, pattern formation, ocular dominance, orientation preference,
pinwheels}
\begin{abstract}
We present an analytical approach for studying the coupled development of ocular dominance (OD)
and orientation preference (OP) columns. Using this approach we demonstrate 
that OD segregation can induce the stabilization and even the production of pinwheels by their
crystallization in two types of periodic lattices.
Pinwheel crystallization depends on the overall dominance of one eye over the other,
a condition that is fulfilled during early cortical development.
Increasing the strength of inter-map coupling induces a transition from pinwheel-free
stripe solutions to intermediate and high pinwheel density states.
\end{abstract}
\pacs{87.19.L-, 05.65.+b, 42.66.Si, 47.54.-r}
\maketitle
\begin{figure}
\includegraphics[width=\linewidth]{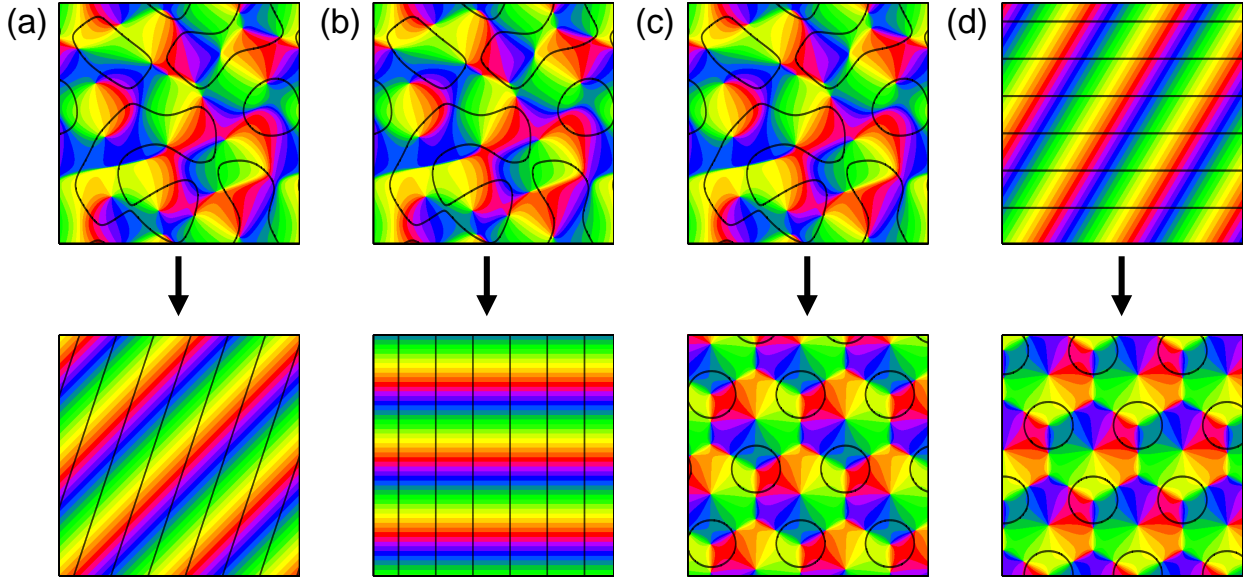}
\caption{
Pinwheel annihilation and preservation in simulations of \refeqn{eq:coupledSpecific} for different strengths of
inter-map coupling and OD bias, $r_o=0.2, r_z=0.02$.
Color coded OP map, zero contours of OD map superimposed. 
(\textbf{\sffamily a}, \textbf{\sffamily b}) $\gamma=0$,$\epsilon=0, 2000$
(\textbf{\sffamily c}, \textbf{\sffamily d}) $\gamma=0.15$,$\epsilon=2000$.
Upper (lower) row: $t=0$ ($10^4/r_z$).
Initial conditions identical in (\textbf{\sffamily a})-(\textbf{\sffamily c}). 
}
\label{fig:Numerics}
\end{figure}
\begin{figure}[t]
\includegraphics[width=\linewidth]{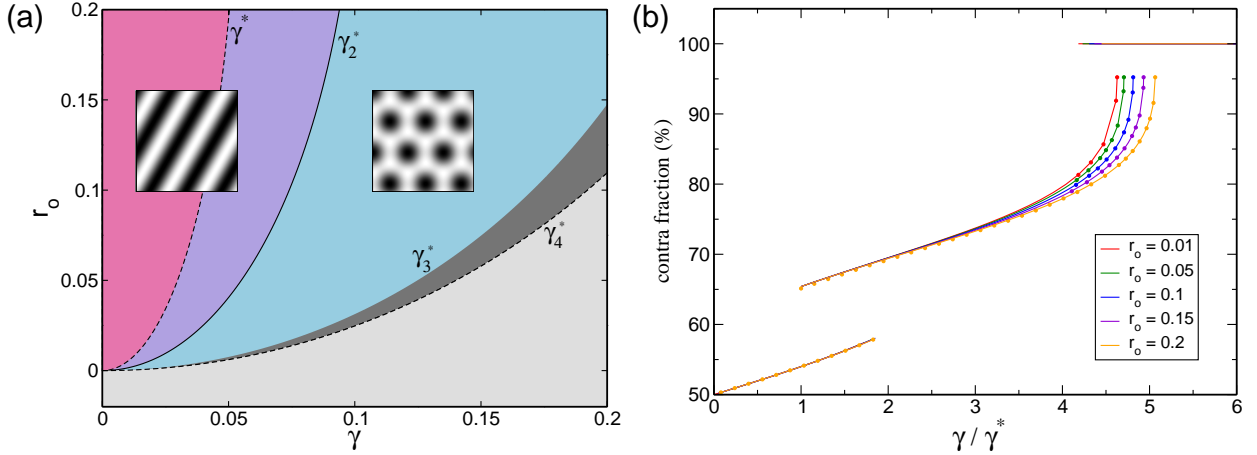}
\caption{
\textbf{(a)} Phase diagram of the OD model \refeqn{eq:coupledSpecific}. Dashed lines:
stability border of hexagon solutions, 
solid line: stability border of stripe solution,
gray regions: stability region of homogeneous solution 
\textbf{(b)} Contralateral eye dominated neurons for the three stationary solutions. 
Circles: numerically obtained values, solid lines: $C_{st}$ and $C_{hex}$. 
}
\label{fig:constTerm}
\end{figure}
\begin{figure}[tb]
\includegraphics[width=\linewidth]{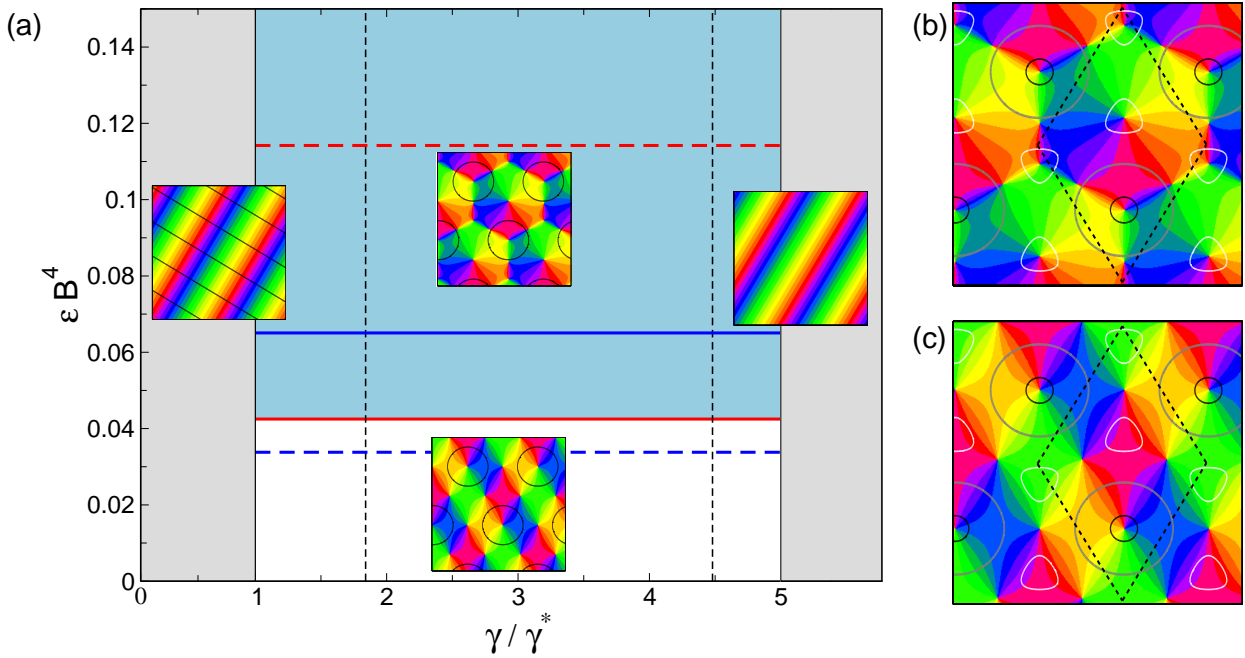}
\caption{(\textbf{a}) Phase diagram of the model \refeqn{eq:AmplEq} for $r_z \ll r_o$.
Vertical lines: stability range of OD hexagons,
red solid line: stability border of hPWC, 
blue solid line: stability border of rPWC,
blue dashed line: transition from rhombic to distorted rPWC.
hPWC is ground state above the dashed red line.
(\textbf{b}) hPWC  (\textbf{c}) rPWC.
OD contour lines: 10\%, 50\%, 90\% contralateral eye dominance (black, gray, white), 
dashed line: unit cell. 
}
\label{fig:PDcoupled}
\end{figure}
\begin{figure}[tb]
\centering
\includegraphics[width=\linewidth]{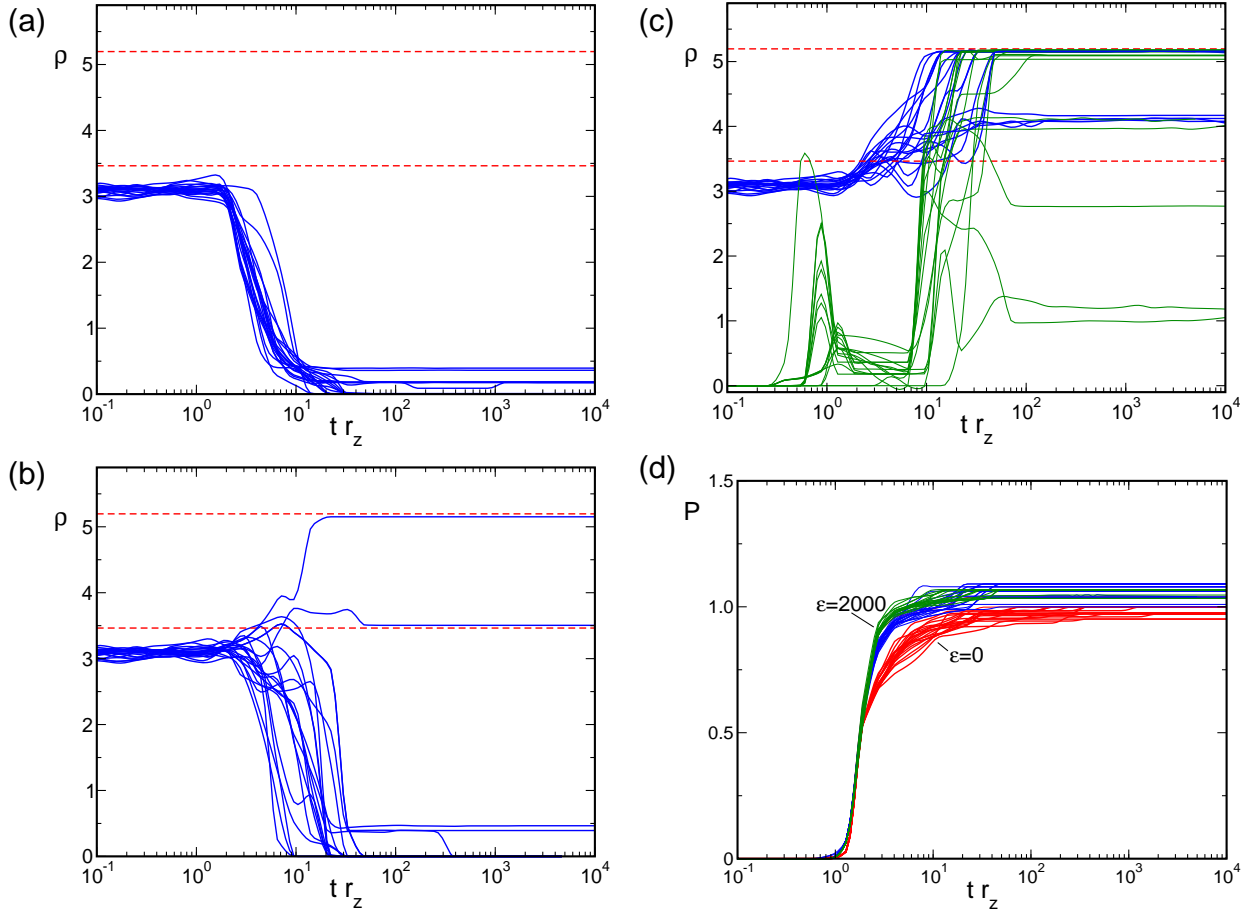}
\caption{Time evolution of the pinwheel density for $r_z=0.05, r_o=0.25, \gamma=0.15$.
(\textbf{a})-(\textbf{c}) Simulations in blue started from an identical set of $20$ initial conditions.
Red dashed lines: $\rho=4\cos(\pi/6)$ and
$\rho=6\cos(\pi/6)$. (\textbf{a})-(\textbf{c}) $\epsilon=0, 200, 2000$.
(\textbf{c}) OD and OP stripes as initial conditions (green).
(\textbf{d}) Power of OP map, $\epsilon=0,200,2000$ (red, blue, green). 
}
\label{fig:PWkin} 
\end{figure}

In the primary visual cortex information is processed in a two dimensional array of
modules called orientation preference (OP) columns \cite{HubelWiesel2}. 
In many species columnar patterns contain pinwheel centers, singular points around which columns activated by
different stimulus orientations are radially arranged like the spokes of a wheel \cite{Swindale2}.
Recent research applying \textit{in vivo} 2-photon imaging
to pinwheel centers revealed that their
radial organization is laid down with single cell precision \cite{PWdetail}. 
How pinwheels are formed during visual development remains unresolved. 
In theoretical models pinwheels are generated by spontaneous symmetry breaking but are often dynamically unstable \cite{Wolf3}.
Recent theoretical studies, 
treating the system of orientation columns essentially as an \textit{isolated} system,
have examined if pinwheels may be stabilized by long-range intracortical 
interactions \cite{Wolf1}, by a coupling to the large scale map of visual space \cite{Retinotopy},
or by wiring length constraints \cite{Koulakov}.
In the visual cortex, however, orientation columns are presumably \textit{interacting}
with e.g. ocular dominance (OD) domains, spatial frequency, and direction
preference columns, see however \cite{Basole}.
For instance OD borders intersect many of these preferentially at right angles 
\cite{Intersection}.
It may thus be inadequate to theoretically study the layout of orientation columns neglecting their
relation to other columnar systems. Recently this perspective has received experimental support by a study
reporting that orientation columns are organized more smoothly when the system of OD columns
is removed \cite{Sur2}.
Indeed simulations suggest that OD segregation impedes the process of
pinwheel annihilation \cite{Wolf3,Cho2}.
So far, however, 
there has been no analytic demonstration that
an intrinsically unstable system
of orientation pinwheels can be stabilized by interactions with other maps. \\
Here we present a dynamical systems approach for analyzing the interactions of
OP and OD maps.
We design a dynamical model for the coordinated development of OP and OD maps
in which pinwheels become unstable in the weak coupling limit.
The inter-map coupling is specified according to experimentally observed geometric
relationships between OD and OP maps.
Because the contralateral eye dominates during the initial formation of OD columns \cite{ODbias},
we systematically study the impact of overall dominance by one eye on the dynamics of pinwheels.
Using weakly nonlinear analysis we derive amplitude equations describing
the existence and stability of pinwheel free and pinwheel rich OP maps in the coupled system.
We  identify two types of pinwheel rich solutions differing in their pinwheel density
and calculate the stability and phase diagram of these solutions
as a function of inter-map coupling and contralateral eye dominance. 
We find that pinwheel crystals are stable above a critical degree of contralateral eye dominance
that induces a patchy pattern of OD domains.
Increasing the strength of inter-map coupling induces a transition from pinwheel free
solutions to low and high pinwheel density patterns. In the latter regime OD segregation even induces
the formation of additional pinwheels. \\
The spatial structure of an OP map is conveniently represented
by a complex field $z(\mathbf{x})$ where $\mathbf{x}$ denotes the 2D position of neurons
in the visual cortex, the modulus $|z(\mathbf{x})|$ is a measure of their selectivity
and $\theta(\mathbf{x})=\frac{1}{2}\arg z$ is their preferred orientation \cite{Wolf3}.
In this representation pinwheel centers are the zeros of the field $z(\mathbf{x})$.
Ocular dominance is described by a real field $o(\mathbf{x})$ where negative and positive
values indicate ipsi- and contralateral eye dominance, respectively.
Because OD and OP maps are not independent of each other
we  consider models containing coupling terms between both fields
\begin{eqnarray}\label{eq:coupledGeneral}
\partial_t \, z(\mathbf{x},t) &=& F[z(\mathbf{x},t),o(\mathbf{x},t)] \nonumber \\
\partial_t \, o(\mathbf{x},t) &=& G[z(\mathbf{x},t),o(\mathbf{x},t)], 
\end{eqnarray}
where $F[z,o]$ and $G[z,o]$ are nonlinear operators.
Various biologically detailed models have been cast in this form \cite{Wolf3,Scherf}.
Because cortical maps arise from a cellular instability with a typical wavelength $\Lambda$,
the mathematically simplest models for the spontaneous generation of these patterns are of Swift-Hohenberg type \cite{Cross}.
We therefore choose $F$ and $G$ to be of this type and couple the fields through
an energy density $T$ 
\begin{eqnarray}\label{eq:coupledSpecific}
\partial_t \, z(\mathbf{x},t) &=& L_z \, z(\mathbf{x},t) -|z|^2z 
-\epsilon \,\frac{\delta T}{\delta \overline{z}} \nonumber \\
\partial_t \, o (\mathbf{x},t) &=& L_o \, o(\mathbf{x},t)-o(\mathbf{x},t)^3+\gamma
-\epsilon \,\frac{\delta T}{\delta o}\, .
\end{eqnarray}
Here $L_{\{o,z\}}=r_{\{o,z\}} -\left(k_{c}^2+\Delta \right)^2$, $\gamma$ is an
OD bias leading to an overrepresentation of the contralateral eye for $\gamma>0$,
and $\epsilon$ is the coupling strength. 
In this model pinwheels are unstable in the weak coupling limit leading to systems
of stripes for $\epsilon=0$, mimicking the behavior of competitive Hebbian models
for OD or OP maps in this situation \cite{Wolf3}.
The form of $T$ is found from the experimental observation that iso-orientation
lines tend to intersect the OD borders perpendicularly \cite{Intersection}.
$T$ can thus be expected to contain terms of the form $|\nabla o \nabla \theta|^{2m}$.
Decomposing the complex field $z(\mathbf{x})$ into the selectivity $|z|$ and 
the preferred orientation $\theta$ finds
\begin{eqnarray}\label{eq:Energy}
T&=&|\nabla z \nabla o|^{2m}=|z|^{2m} \left( |\nabla o \nabla \ln |z| |^2+4|\nabla o\nabla \theta|^2  \right)^m\nonumber \\
&=& \left( |\nabla o \, \nabla \textrm{Re} z|^2 + |\nabla o \, \nabla \textrm{Im} z|^2 \right)^m .
\end{eqnarray}
Where orientation selectivity is locally homogeneous, i.e. $ \nabla \ln |z| \approx 0$,
$T$ is minimized if the direction of the iso-orientation lines ($\nabla \theta$) is
perpendicular to the OD borders. 
At pinwheel centers the zero contours of $\textrm{Re}\, z$ and $\textrm{Im} \, z$ cross
and $\nabla \textrm{Re} \, z$ and $\nabla \textrm{Im} \, z$ are not parallel, 
$T$ can be minimized only if $|\nabla o|$
is small at the pinwheel centers, i.e. near extrema or saddle-points of $o(\mathbf{x})$.
In the following we analyze the case $m=2$. As we will see below, this choice allows 
for a limit in which map interactions become unidirectional.\\
We observe that for substantial contralateral bias and above a
critical coupling $\epsilon$ pinwheels are preserved or
are even generated after symmetry breaking.
Numerical simulations of the dynamics \refeqn{eq:coupledSpecific} are shown in 
\reffig{fig:Numerics}.
Without a contralateral bias the attractors are pinwheel-free stripe solutions 
irrespective of the strength of the inter-map coupling. \newline
To reveal the exact conditions for the preservation of pinwheels by inter-map coupling
we used weakly nonlinear analysis to study the nature and stability of different types of solutions.
To this end we first studied how the emerging OD map depended on the overall eye dominance.
Shifting the OD field by a constant $o(\mathbf{x},t)=\tilde{o}(\mathbf{x},t)+\delta$,
the dynamics \refeqn{eq:coupledSpecific} is mapped to
$\partial_t \, \tilde{o}(\mathbf{x},t)=\tilde{L} \, \tilde{o}+\tilde{\gamma}\tilde{o}^2
-\tilde{o}^3$ with $\tilde{L}=\tilde{r}_o-\left( k_{c}^2+\Delta\right)^2$, $\tilde{r}_o=r_o-3\delta^2$, and $ \tilde{\gamma}=-3\delta$ where $\delta$ is the real solution of $-\delta^3+\left(r_o-k_c \right) \delta +\gamma=0$;
an equation that has been extensively studied in pattern formation literature \cite{Soward}. 
It has three types of stationary solutions:
(1) A homogeneous solution with spatially constant eye dominance $o_c(\mathbf{x})=\delta$,
(2) OD stripes $o_{st}(\mathbf{x})=2\mathcal{B}_{st}\cos \left(x+\psi \right) +\delta,$
with $\mathcal{B}_{st}=\sqrt{\tilde{r}/3}$, and (3)
hexagonal arrays of ipsilateral eye dominance blobs in a sea of contralateral eye dominance 
$o_{hex}(\mathbf{x})=\mathcal{B}_{hex}\sum_{j=1}^3 e^{\imath \psi_j}e^{\imath \vec{k}_j \vec{x}}+c.c. +\delta, $
with $\sum_j^3 \vec{k}_j=0$, $\sum_j^3 \psi_j=\pi$, 
and $\mathcal{B}_{hex}=-\tilde{\gamma}/15+\sqrt{\left(\tilde{\gamma}/15\right)^2+\tilde{r}/15}$.
The fractions of contralateral eye dominated territory $C_{st}$ and $C_{hex}$ increase with $\gamma$ as 
$\cos \left( C_{\textrm{st}}\pi \right) = -\delta/\left(2\mathcal{B}_{st} \right)$ 
and $(1-C_{\textrm{hex}})\sqrt{3}\, 2\pi \approx 3\, \arccos \left(\frac{1}{2}\left(-1+\sqrt{3+\frac{\delta}{\mathcal{B}_{hex}}}\right)\right)^2$ for (2) or (3),
\reffigPouter{fig:constTerm}{b}. 
The phase diagram of this model is depicted in \reffigP{fig:constTerm}{a}.
It shows the stability borders for the three solutions.
Without a bias term the OD map is either constant, for $r_o<0$, or has a stripe layout, for $r_o>0$.
For positive $r_o$ and increasing bias term there are two transition regions. First a transition
region from stripes to hexagons (between $\gamma^*$ and $\gamma_2^*$) and second a transition region from hexagons to the homogeneous solution (between $\gamma_3^*$ and $\gamma_4^*$).
\newline
Close to instability, stationary solutions
to the full dynamics \refeqn{eq:coupledSpecific} can be calculated analytically
by weakly nonlinear analysis \cite{Manneville}. The Fourier components
of the emerging pattern are located on the critical circle 
$\vec{k}_j=\left( \cos j\pi/3 , \sin j\pi/3 \right) k_{c}$ so that
\begin{eqnarray}\label{eq:Modes}
z(\mathbf{x},t)&=&\sum\limits_j^{3} \left( A_j(t)e^{\imath  \vec{k}_j \vec{x}} + A_{j^-} (t)  e^{-\imath \vec{k}_j \vec{x}} \right)\nonumber \\
o(\mathbf{x},t)&=&\sum\limits_j^{3} \left( B_j(t)e^{\imath \vec{k}_j \vec{x}} + \overline{B}_{j} (t) e^{-\imath  \vec{k}_j \vec{x}} \right) ,
\end{eqnarray}
with the complex amplitudes $A_j=\mathcal{A}_j e^{\imath \phi_j}$,  
$B_j=\mathcal{B}_j e^{\imath \psi_j}$.
Although the coupling terms enter at seventh order in the amplitude expansion they can be written
as an effective cubic interaction term.
Because $A_i \propto \sqrt{r_z}$ and $B_i \propto \sqrt{r_o}$, the coupling onto the OD dynamics becomes small for $r_z \ll r_o$, 
since terms like $\epsilon |A|^4|B|^2B\propto r_z^2 r_o^{3/2}$ are negligible compared to terms like $|B|^2B\propto r_o^{3/2}$.
In this limit, the backreaction of the OP map onto the OD map is thus negligible.
Using uniform modes $\mathcal{B}_i=\mathcal{B}$, the amplitude equations for the OP map
are given by
\begin{eqnarray}\label{eq:AmplEq}
\partial_t \, A_i  &=& r_z\, A_i -\sum\limits_j^6 g_{ij}|A_j|^2A_i -2\sum\limits_{j\neq i} ^3 A_j A_{j^-}\overline{A}_{i^-} \nonumber \\
&&  -\epsilon\,  \mathcal{B}^4 \sum_{j,l,k}^6  h_{ijlk} A_j A_l \overline{A}_k \, ,
\end{eqnarray}
with $A_{j^-}=A_{j+3}$, $g_{ii}=1$, $g_{ij}=2$ and $h_{ijlk}$ an effective self-interaction tensor.
The dynamics of the modes $A_{i^-}$ is given by interchanging $A_i$ and $A_{i^-}$.
A solution of hexagonal symmetry (symmetric under rotation by 120 degree) to \refeqn{eq:AmplEq} is given by the uniform solution
$\mathcal{A}_j = \mathcal{A}_{j^-}=\mathcal{A}$, $\phi_j = \psi_j+(j-1)2\pi/3+d\, \delta_{j,2}$,
and $\phi_{j^-}=-\psi_j+(j-1)2\pi/3+d\, \left( \delta_{j,1}+\delta_{j,3}\right)$,
where we choose $\psi_1=\psi_3=0,\psi_2=\pi$ and the constant 
$d\approx 1.176$ is the solution of a transcendental equation.
For negligible backreaction $\mathcal{B}=\mathcal{B}_{hex}$
and $\mathcal{A}^2\approx r_z/ \left(9+55.6 \epsilon \,  \mathcal{B}_{hex}^4  \right)$.
The uniform solution is determined up to a free phase $\varphi$ which results from the orientation
shift symmetry $z\rightarrow z \, e^{\imath \varphi}$ of \refeqn{eq:coupledSpecific}. The positions of the pinwheels are fixed by the OD map and there are no translational degrees of freedom.
In addition to these hexagonal pinwheel crystals (hPWCs) there exist also non-uniform solutions.
Besides stripe-like solutions of $z(\mathbf{x})$ with one dominant mode
we find rhombic pinwheel crystals (rPWCs) $\mathcal{A}_j=\mathcal{A}_{j^-}=\left(\mathcal{A},a,\mathcal{A}\right)$
with $a \ll \mathcal{A}$ and distorted rhombic crystals 
$\mathcal{A}_j=\left(\mathcal{A}_1,\mathcal{A}_2,\mathcal{A}_3\right), A_{j^-}=\left(\mathcal{A}_3,\mathcal{A}_2,\mathcal{A}_1\right)$, both
symmetric under rotation by 180 degree. 
We analytically calculated the stability properties of the uniform solution by linear stability analysis.
The phase diagram for $r_z\ll r_o$, cf. \reffigP{fig:PDcoupled}{a} reveals a transition
from rPWCs to hPWCs with increasing coupling strength $\epsilon$
for intermediate degrees of OD bias. 
For $\gamma<\gamma^*$ or for $\gamma>\gamma_4^*$ pinwheel-free orientation stripes
are dynamically selected. For $\gamma^*<\gamma<\gamma_4^*$ and above a critical effective coupling
strength $\epsilon \mathcal{B}^4\approx 0.042$, hPWCs are stable and become 
the energetic ground state above $\epsilon \mathcal{B}^4 \approx 0.12$.
Below $\epsilon \mathcal{B}^4 \approx 0.065$, rPWCs are stable leading to a bistability
region between rPWCs and hPWCs.
We find in this region that rPWCs transform into distorted rPWCs above an effective
coupling strength of  $\epsilon \mathcal{B}^4 \approx 0.033$.
Although rPWC are stable even in the uncoupled case they never become the energetic ground
state.
Thus for substantial bias towards one eye pinwheels are in fact stabilized and pinwheel
rich solutions become ground states by inter-map coupling.\newline
The layouts of the main pinwheel rich solutions are shown in \reffigPP{fig:PDcoupled}{b,c}.
The hPWC contains 6 pinwheels per unit cell and 
the pinwheel density i.e. the number of pinwheels per $\Lambda^2$ \cite{Wolf3}
is given by $\rho=6 \, \cos (\pi/6) \approx 5.2$.
The rPWC has 4 pinwheels per unit cell and its pinwheel density is
$\rho=4 \, \cos (\pi/6) \approx  3.5$.
One may expect that the energy term \refeqn{eq:Energy} favors pinwheels to colocalize with OD extrema.
For hPWCs three pinwheels of the same topological charge are in fact
located at the extrema of the OD map.
The other three however are located near OD borders. 
In case of the rhombic layout there is only one pinwheel at an OD extremum while the
other three pinwheels are located at OD saddle-points which are also energetically favorable
positions with respect to $T$. \newline
We tested whether these solutions and their stability ranges revealed for $r_z \ll r_o$
persist when the backreaction on the OD map is taken into account. To this end we
solved 
the full field dynamics \refeqn{eq:coupledSpecific} numerically using a fully implicit Krylov subspace algorithm 
with periodic boundary conditions on a 128x128 mesh with an aspect ratio of
$\Gamma=22$.
In simulations we tracked the pinwheel density from $t=0$ to $t=10^4 \, r_z^{-1}$, cf. \reffig{fig:PWkin}.
In the uncoupled case ($\epsilon=0$), most of the patterns decay into a stripe solution
and their pinwheel density drops towards zero. At small coupling strengths ($\epsilon=200$) 
the pinwheel density
converges either to zero (stripes), to values near 3.5 for the rPWC, or
to approximate 5.2 for the hPWC. At high map coupling ($\epsilon=2000$), 
pinwheel-free stripe patterns form neither from pinwheel rich nor from pinwheel free initial conditions.
In this regime the dominant layout is the hPWC. However regions of hPWC layout
can be inter-digitated with long lived rPWCs and stripe domains.
\reffigBEGINP{fig:PWkin}{d} shows the time course of the power $P(t)=\langle |z(\mathbf{x},t)_{dyn}|^2 \rangle _\mathbf{x} / \langle |z(\mathbf{x},t)_{th}|^2 \rangle _\mathbf{x}$. The field $z_{th}$ is obtained from solution of the amplitude equations \refeqn{eq:Modes} while $z_{dyn}$ is the field obtained from the simulation.
The amplitudes grow and saturate after $t \approx r_z^{-1}$. 
When the amplitudes are saturated pattern selection starts.
Quantitatively, we find that with backreaction the critical coupling strengths are slightly
increased compared to their values in the limit $r_z \ll r_o$. 
\newline
Our analysis for the first time conclusively demonstrates that OD segregation can stabilize pinwheels,
even if they are intrinsically unstable in the uncoupled dynamics of the OP map,
raising the possibility that inter-map coupling is the mechanism of pinwheel stabilization
in the visual cortex.
Our results indicate that the overall dominance of one eye is important for the
effectiveness of this mechanism.
In this case, OD domains form a system of patches rather than stripes enabling
the capture and stabilization of pinwheels by inter-map coupling.
Studying a wide range of phenomenologically conceivable interaction energies we find that
systems of OD stripes are in general not expected to stabilize pinwheel patterns. 
Interestingly, visual cortex around the time of early OP development
is indeed dominated by one eye and has a pronounced patchy layout of OD domains supporting this notion \cite{ODbias}.
Further support comes from experiments in which the OD map
was removed artificially, resulting in a significantly smoother OP map \cite{Sur2}. Removal
of the OD map, however, does apparently not completely destabilize pinwheels.
This might reflect the influence of additional columnar systems of patchy layout like spatial frequency or 
direction columns that are expected to interact with the OP map in a similar fashion
as OD columns. Interactions among multiple coupled maps may potentially
also explain the non-crystalline spatial organization of OP maps in the visual cortex.
The approach introduced here will be useful for further rigorous analyses of the interaction
among multiple maps in cortical development.
\acknowledgments{We thank D. Heide and M. Kaschube for providing the Krylov algorithm and for fruitful discussions.
We thank M. Huang for discussions.
This work was supported by the HFSP and BMBF.}

\bibliographystyle{apsrev}

\end{document}